# Multiband Tunable Large Area Hot Carrier Plasmonic-Crystal Photodetectors

F. Pelayo García de Arquer[†], Agustín Mihi[†,*], Gerasimos Konstantatos*

* gerasimos.konstantatos@icfo.es, agustin.mihi@icfo.es
† - equal contribution

ICFO - Institut de Ciències Fotòniques, Mediterranean Technology Park 08860 Castelldefels, Barcelona, Spain

**Optoelectronic functionalities of photodection and light harnessing rely on the band-to-band excitation of semiconductors, thus the spectral response of the devices is dictated and limited by their bandgap. A novel approach, free from this restriction, is to harvest the energetic electrons generated by the relaxation of a plasmonic resonance in the vicinity of a metal-semiconductor junction. In this configuration, the optoelectronic and spectral response of the detectors can be designed ad hoc just by tailoring the topology of metal structures, which has tremendous applications in solar energy harvesting and photodetection. Fully exploiting hot electron based optoelectronics yet requires a platform that combines their exotic spectral capabilities with large scale manufacturing and high performance. Herein we report the first implementation of a large area, low cost quasi 3D plasmonic crystal (PC) for hot electron photodetection, showcasing multiband selectivity in the VIS-NIR and unprecedented responsivity of 70 mA/W.**

The ability to fabricate hot carrier based devices with tailored spectral response, is exciting for solar energy harvesting,[1–6] and visible or infrared detection.[4–15] These devices have relied on nanoparticles,[8] nanoantennas[10,15] or gratings[11] to excite either localized or surface plasmons whose relaxation will produce the hot carriers. Such metallic nanostructures are fabricated with costly and time-consuming lithographic processes thus limiting their potential for large-area high-throughput manufacturing required for realistic applications. Here we report the implementation of a large area, low cost plasmonic crystal (PC) as a novel platform for hot electron generation. The PC facilitates the excitation of a wide variety of propagating, localized and hybrid plasmon resonances, which ultimately decay in the hot carriers exploited herein. A tunable and multiband VIS-NIR photodetector with responsivities of 70 mA/W under bias, is fabricated via soft nano-imprinting lithography (NIL),[16] paving the way for the implementation of this technology at a large-scale for a vast variety of optoelectronic devices.

To excite plasmon resonances in a metal, one has to account for the momentum mismatch between the bound surface plasmon and the free propagating incoming photon.[17] This additional momentum can be provided, for instance, by scattering of nanoparticles, a prism or a diffraction grating. The latter configuration is advantageous since the periodicity of the architecture determines the frequencies at which surface plasmon polaritons (SPPs) are excited, irrespective of the plasmon frequency of the metal.[18,19] Furthermore, if the metal is periodically corrugated or structured, surface plasmons can interfere forming standing waves. This is the case of the Plasmonic Crystals (PCs), in reference to their dielectric counterparts (Photonic Crystals), metallic architectures with periodically organized motifs.[20] Plasmonic crystals can sustain a wide variety of plasmonic resonances (propagating and localized) that can further couple with each other to produce strong hybrid resonances.[21,22] In this manuscript, we intend to exploit the complex and design-dependent dispersion relation of the plasmonic crystal architecture to fabricate a tunable and broad band multispectral VIS-NIR hot-electron based large area photodetector.

**Results**

Our plasmonic crystal based photodetector is represented in Figure 1. Briefly, a 2D square array of cylinders is shaped in a photoresist via soft nano-imprinting using a pre-patterned PDMS (Polydimethyl-siloxane) mold; a scalable-large area technique compatible with roll to roll manufacturing.[16] Next, an ITO layer is sputtered on top (bottom electrode), followed by a conformal $TiO_2$ coating passivated with an ultrathin layer of alumina,[4] upon which a gold metal (top electrode) is deposited (see Fig. 1b.). Following this approach, 9 mm$^2$ plasmonic crystal based photodetectors of with different geometric characteristics were fabricated (see photograph in Fig. 1c.). The bright colors observable with the naked eye are a result of the periodic single crystalline metallic corrugation of the structure offered by NIL. The optical properties of the resulting plasmonic crystals can be precisely tuned with the lattice parameter (*L*), radius (*r*), cylinder depth or TiO$_2$ thickness. A SEM image (45º angle) of the metal film (Fig. 1d.) illustrates the periodic arrangement of the plasmonic crystal. The constituting ITO, TiO$_2$ and Au layers of the architecture can be distinguished in the cross-sectional SEM image in Fig. 1e. The optoelectronic operation of our photodetector builds on the interaction of the corrugated metal film with light. Photons impinging from the ITO side (Fig. 1a.) excite several plasmonic resonant modes in the metal, which ultimately result in the creation of electron-hole pairs in the metal by Landau damping (Fig. 1f.). This rising hot electron population can then be collected by the Schottky barrier built at the Au-TiO$_2$ interface, resulting in a photocurrent ultimately dictated by the plasmonic crystal resonances.[10]

Figure 2 illustrates the spectral tunability of the responsivity at short-circuit conditions of our photodetectors varying the plasmonic crystal geometry. The changes produced in the excited plasmonic resonances depend strongly on the lattice parameter, radius or TiO$_2$ coating thickness, as evidenced by the associated responsivities derived from hot-carrier emission. Using nano-imprinting stamps with predefined features, photodetectors with different lattice parameter *L* (450 nm and 550 nm) and/or radius *r* (200 nm and 150 nm) were built. The degree of control over this platform is manifold: the number of resonances, their intensities and positions depend strongly on these parameters. Just modifying the pillar radius by 50 nm (Fig. 2a and b) results in a prominent increase of the intensity of the NIR resonances at λ = 750 nm and λ = 900 nm. A broadband spectrum is obtained from the overlap of a set of narrowband resonances. This is, to our knowledge, the first demonstration of a solid state plasmonic photodetector with such multispectral resolution at such a large scale. Flat reference responsivity is also shown for comparison in Fig. S1.

To gain insight into the underlying mechanism that gives rise to this multiband response, the optical absorption of a nanostructured device ($r = 150$nm, $L = 450$nm and 60nm of $TiO_2$) was simulated and compared with its experimental responsivity. A set of resonances that extend from the visible to the near infrared peaking at 515 nm, 714 nm and 886 nm can be observed both in the experimental responsivity and in the simulated absorption in the gold film (Fig. 3a). The physical origin of these resonances can be inferred by calculating the electric field intensities (which indicate where light is concentrated in the structure, Fig. 3b-c, top) and the absorption rate profiles (to expose only the regions where light is absorbed, Fig. 3b-c bottom) along a device cross section for the three wavelengths indicated with vertical dashed lines in Fig. 3a.

The field intensity profile for $\lambda_1 = 515$ nm (Fig 3b, top) spreads within the $TiO_2$ layer, the photoresist and penetrates into the metal plate and cylinder. The corresponding absorption is found within the volume of the gold pillar (Fig. 3b, bottom). The electric field profile in this case suggests a complex interplay between the plasmons excited by the grating periodicity and the bulk absorption of the Au pillars (the interband transition in gold starts at 2.4eV).[22] In this frequency range, light can penetrate within the gold cylinder (an architecture with r = 150nm has an Au cylinder with 50 nm radius, considering the ITO and $TiO_2$ coatings), allowing the excitation of lossy modes inside the cylinder. The divergence of simulated absorption and experimental responsivity bellow 450 nm can be ascribed to a less efficient hot-electron injection from the pillar-core compared to its surface,[23,24] as the electron mean free path is of the order of 15-20 nm for these wavelengths. Gold interband transitions are also known to be less efficient in hot-electron injection due to the depth of Au *d*-band levels.[25]

The longer wavelength resonance at $\lambda_2 = 714$ nm (Fig. 3c.) presents on the other hand an electric field distribution strongly bound to the Au/TiO$_2$ interface. High intensity nodes are identified at the base and at the top of the metal pillar, suggestive of SPP interference leading to very high localization. For $\lambda_3 = 886$ nm the field is more concentrated along the vertical facets of the metal cylinder. It is important to note that the corresponding spatial distribution of the absorbed photons is more intense at the interface with the titanium dioxide for the latter two plasmonic resonances, thus facilitating the injection of the hot electrons into the oxide due to the proximity with the interface.

A deeper insight in the physical origin of the three plasmonic modes described before can be obtained by comparing the evolution of the 2D absorption map (wavelength versus lattice parameter *L*) when altering the components from the original architecture (Figure 4). These 2D maps help identify the plasmonic modes that depend only on the diffraction condition imposed by the lattice parameter (*L*), therefore called Bragg – SPP modes, the ones that are *L*-invariant, and the regions where both modes are coupled.[21,22,26] Several of these 2D diagrams are analyzed in what follows. First, the effect of the ITO absorption towards longer wavelengths is revealed when the 2D absorption map is computed considering only the real part of the refractive index of the conductive oxide (Fig. 4a-b.). When the losses in the ITO are disregarded, the optical response of the architecture extends deep into de NIR with increasing lattice parameter. Second, a simplified architecture is considered, removing the glass, resist and ITO layers, leaving a 2D titania – gold corrugated interface (Fig 4c). In this latter simulation, the number of plasmonic modes is dramatically reduced, implying that many of the modes are excited from the multilayered grating architecture that constitutes our plasmonic crystal. The different corrugated interfaces (resist-ITO, ITO-TiO$_2$, TiO$_2$-Au) effectively provide with several subsequent gratings of similar center to center distance but different radii and refractive indexes, launching different SPPs.

Finally, we intend to distinguish between the Horizontal, *L*-independent bands observed together with the clear Bragg-SPPs. The first correspond to localized plasmons and will depend on the gold nano-pillar geometry (Fig. S2-S3). The latter are associated with delocalized propagative waves along the metal-semiconductor interface. When both modes overlap, strong absorption enhancement and broadening is observed, as is the case for the point (*L*,λ) = (500 nm, 714 nm). The excitation channels derived from a pure photonic grating on a flat metal film are depicted in Fig. 4d, depicting the preferential coupling to Bragg-SPPs modes. The absence of intense resonances underlines the important role of the metal corrugation to observe the strongly localized modes described previously.[27] The light incoupling properties of our PC can be understood in a simplified picture as the interaction of individual metallic (Fig 4b) and dielectric (Fig. 4d) gratings. The mixing between localized and Bragg plasmons has been reported to greatly modify the mode distribution and enhance their oscillator strength.[21,22] Such hybridization is evident now in Fig. 4b, where intense propagative Bragg –SPP modes dominate above 800 nm, and the coupling of those with the localized modes appearing at the edges of the metallic cylinders. The origin of $\lambda_2$ and $\lambda_3$ resonances is attributed therefore to the interaction of Bragg and localized plasmon modes (Fig. 3c) and to propagative modes (Fig. 3d), respectively.

Further structural parameters can be employed as leverage to tune and extend the spectral response of this architecture into longer wavelengths. The variation of the optical response of our architecture with the refractive index and thickness of the oxide layer, as well as with the pillar depth and radius is summarized in Figures S2 and S3.

We now turn to the optoelectronic characteristics of these structures as photodetectors. The dark current − voltage response of our detectors, indicative of a Schottky junction, is shown in Fig. S4. The responsivity of our devices increases by orders of magnitude when the Schottky junction is negatively biased (see Fig. 5). There is a sharp increase in photoconversion efficiency with applied voltage, as the strong $Al_2O_3/TiO_2$ barrier modulation under reverse bias favors the injection of the hot-electron population.[4,15] As the reverse bias is further increased most of the excited population can tunnel and another regime with a milder responsivity-bias dependence appears. A similar behavior was also recently reported in MIM hot-electron devices.[15] Noticeable, both responsivity onset and saturation features depend on the energy of the excitation $h\omega$, as depicted for 406 nm, 640 nm and 900 nm illuminations in Figure 5. The hot electron population will span along $E_f$ and $E_f+h\omega$ (ref. 23), and consequently higher energy excitations are expected to produce responsivity onsets and saturations at lower voltages, which we confirmed experimentally.

In the saturation regime, responsivities as high as 70 mA/W are within reach, being ultimately limited by the absorption of the structure and the hot electrons emission. This responsivity is the highest reported ever for plasmonic hot carrier devices. The corresponding external quantum efficiency (ratio of collected carriers to number of incident photons) in this case is 12%, yielding an estimate (see suppl. section S5) for the internal quantum efficiency up to 30% (ratio of collected carriers to number of absorbed photons). The dynamic range characteristics of our devices are showcased in Fig S6, where a linear response of the photocurrent as a function of incident power is shown. Their temporal response (Fig S7), in the tens of ms range, is currently limited by the *RC* constant of the devices in view of their large area footprint and further improvements can be readily achieved via appropriate geometric designs.

In summary, we present a robust and novel architecture for multispectral visible-infrared photodetection based on hot-electron injection from plasmonic crystals. Our approach is not only interesting from a photonics point of view due to the rich optical response of the plasmonic crystals, but also relies on soft nano-imprinting lithography and is thus compatible with large area manufacturing, a requisite for scalable production. Our devices can yield responsivities up to 70 mA/W and IQEs up to 30% in the presence of an electric field, enabling the use of this technology for practical applications. Improvement in responsivities can be expected by further control of hot electron emission and collection directions, shown to increase the collection efficiency of hot electrons in other systems.[14] Other areas such as photovoltaics or photocatalysis[5,6] can benefit from this approach, as large area and infrared sensitization can result in reduced fabrication costs and increased efficiencies respectively. Further development might rely on specific plasmonic crystal designs and the exploitation of plasmonic non linear processes.[28] This architecture can also serve as a platform for energy harvesting at mid-infrared or THz regions[29] if scaled and applied to metal-insulator-metal systems.[9,15]

**Methods**

**Device fabrication:** Nanostructured electrodes were fabricated by nano-imprinting lithography. Roughly, a poly(dimethylsiloxane) mold (PDMS, 10:1 Sylgard 184) replicating a square array pattern from a silicon wafer (AMO gratings) was used as "printing stamp" on an ethanol wetted 500 nm film of UV photo curable epoxy (SU8, Microchem) spin casted on an ITO coated substrate. The resist is UV cured ($\lambda = 360$ nm, 15 min) through a spatial mask defining squared imprinted regions of 9 mm$^2$. After exposure and development of the resist, 40 nm of ITO are sputtered on the substrates using an AJA sputtering system without masking so the sputtered ITO is electrically connected to a commercial ITO coated substrate. Finally a 90 min anneal at 200°C under inert atmosphere is carried out to improve the conductivity of ITO. Further fabrication details can be found elsewhere.[30]

60-80nm $TiO_2$ and 0.5nm $Al_2O_3$ were deposited by atomic layer deposition (Savannah 200, Cambridge Nanotech).Water and trimethylaluminiumwere used as precursors of $Al_2O_3$ with open valve times of 0.05 and 0.10 s, respectively, followed by 65 s pump time (1.1 A per cycle). The chamber was kept at 150ºC during the deposition process. TiO2 coatings were fabricated from water and titanium isopropoxide; valves were set to 0.015 and 0.065 s, respectively, followed by 10 s pump time (0.025 A per cycle). The deposition chamber temperature was held at 200ºC.

200 nm of metal deposited on a Kurt J. Lesker Nano 36 systemat a rate of 0.5 Å s$^{-1}$ for the first 30 nm and 1.5 Å s$^{-1}$ for the remaining 250 nm, at a base pressure lower than 2 10$^6$ mbar.

**Device characterization:** all device characterization was performed in ambient conditions. Current-voltage characteristics were recorded using a Keithley 2636 source meter. The spectral characterization was carried out by illuminating the devices with by a Newport Cornerstone 260 monochromator, modulated with a 20 Hz chopper, and monitoring the short circuit current with a Stanford Research System SR830. Over 200 spectrum traces were acquired and averaged for each measurement to avoid noise artifacts. Responsivity measurements with bias were taken under a 635 nm laser illumination with a frequency of 100 mHz and a power density of 3 mW/cm$^2$, and the current measured with a Keithley 2636 source meter. Short-pulsed illumination was achieved with a 4-channel Thorlabs laser controlled by an Agilent 33220A waveform generator or alternatively with a NKT Photonics supercontinuum laser source and recorded with an Agilent B1500A semiconductor parameter analizer. The photocurrent-power dependence was acquired at short-circuit conditions with a Stanford Research System SR830 under 635 nm diode laser illumination from Newport (LQA635-08C) controlled by an Agilent 33220A waveform generator.

**Device modeling:** Finite-Difference Time-Domain Simulations (FDTD) were carried out using Lumerical FDTD solutions suite version 8 (http://www.lumerical.com). A unit cell illuminated with a plane wave source was simulated with symmetric and anti-symmetric boundary conditions. The metal electrode refractive index was taken from Johnson and Christy and the absorbing ITO built from absorption measurements using Kramers-Kronig relations.

**Acknowledgements**

We acknowledge financial support from Fundació Privada Cellex Barcelona, the European Commission's Seventh Framework Programme for Research under Contract PIRG06-GA-2009-256355 and the Ministerio de Ciencia e Innovación under Contract No. TEC2011-24744. We also acknowledge support from the Nanophotonics for Energy Network of Excellence under contract N4E GA.248855. A.M was supported by an ICFONEST fellowship (COFUND program). We are grateful to D. Torrent for fruitful discussions.

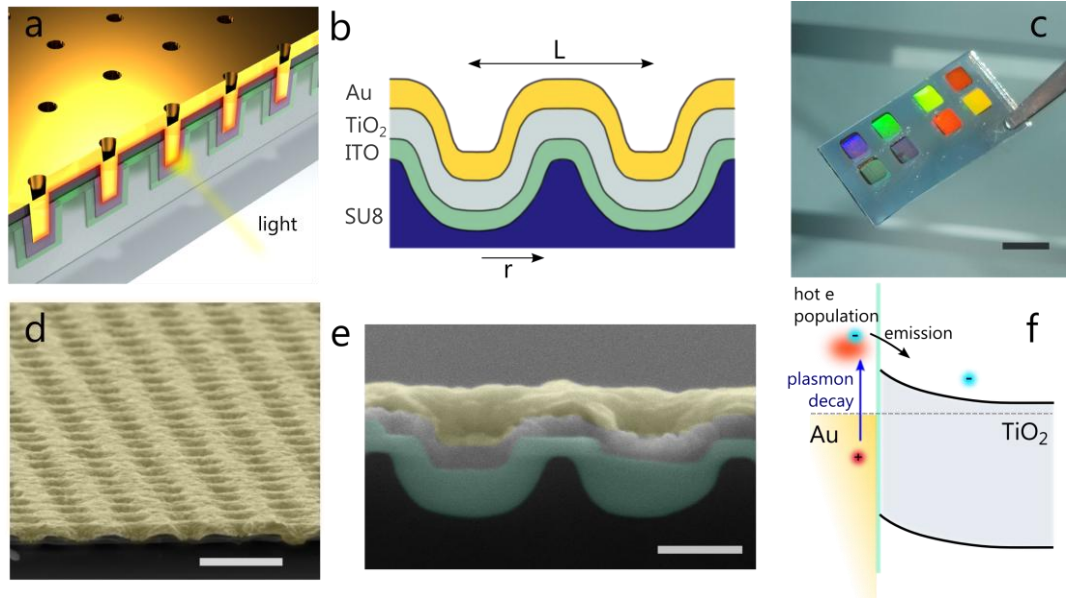

**Figure 1. Device architecture and principle of operation.** (a) Representation of the plasmonic crystal photodetector. Light impinges from the bottom (ITO/glass) exciting resonant modes responsible for the hot electron generation. (b) Schematic of the device architecture: a square array of cylindrical voids in photoresist is coated with 40nm of ITO, followed by 60nm of TiO$_2$ and 150nm of Au. Different geometries are fabricated varying both lattice parameter (*L*) and / or the cylinder radius (*r*). (c) Photograph of a substrate containing eight 9 mm$^2$ devices; the reflected colors are indicative of the nanostructured metal electrodes. Scale bar is 1 cm. (d) 45° angle view SEM image of the periodic arrays. Scale bar is 2 μm. (e) Cross sectional SEM artificially colored to portray the different layers of the architecture. Scale bar 400 nm. (f) Schematic representation of the photocurrent generation process after light excitation: hot electrons derived from plasmonic damping are emitted over the Au/TiO$_2$ Schottky barrier into the TiO$_2$ conduction band.

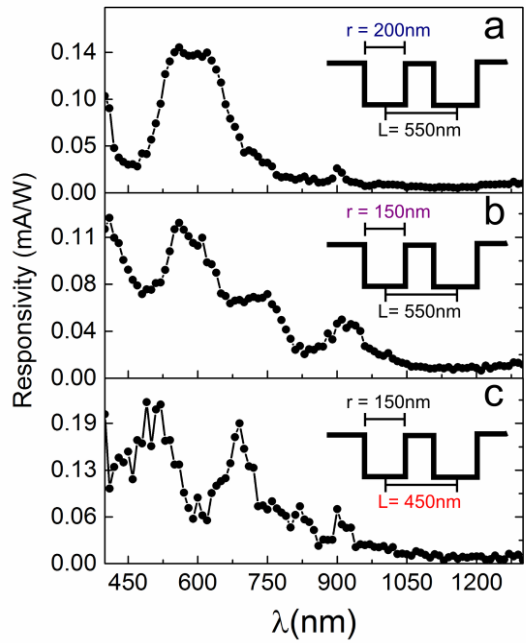

**Figure 2. Experimental responsivity of a plasmonic crystal based photodetectors** for different structural parameters measured at short circuit conditions: (a) r= 200nm and L=550nm; (b) r= 150nm and L=550nm and (c) r= 150nm and L=450nm. The spectral selectivity, number of resonances, their intensities and positions can be modified with these parameters.

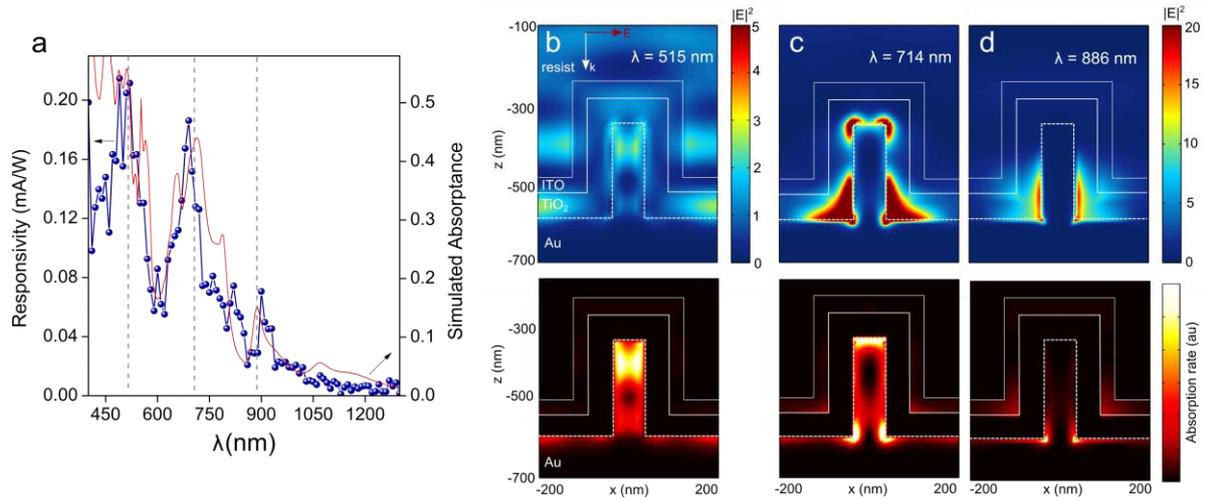

**Figure 3. Spectral Responsivity of the devices and its correlation with the optical response of the plasmonic crystal** (a) Responsivity (scattered points) versus the FDTD simulated absorption in the gold (red line). The electric field distribution (top) and absorption rate (bottom) within the plasmonic architecture at (b) $\lambda_1$ = 515 nm (c) $\lambda_2$ = 714nm and (d) $\lambda_3$ = 886nm illustrating the different nature of the resonances, from Mie-Bragg (b) to Localized-Bragg (c) to Bragg SPP (d) modes.

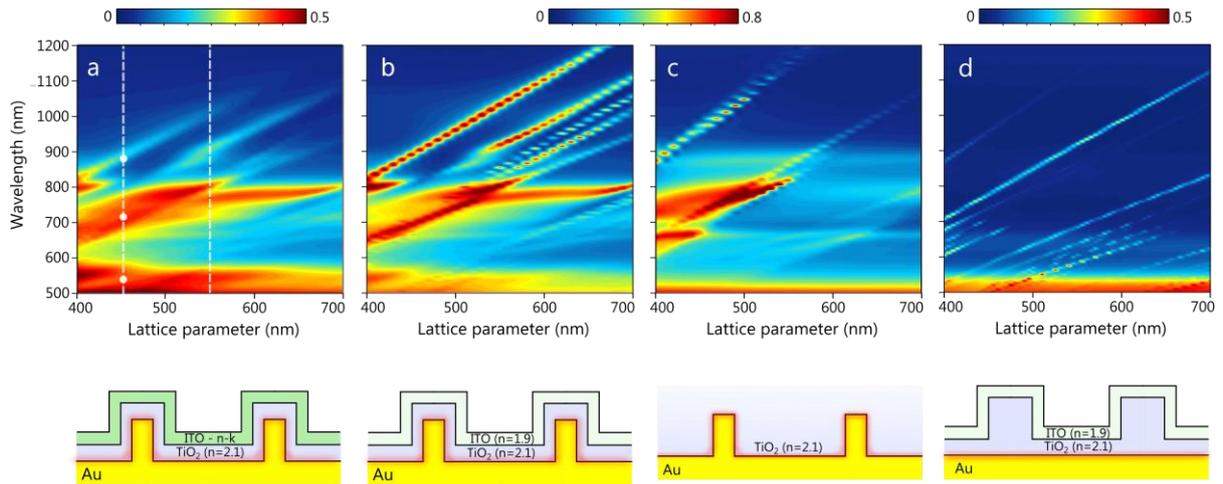

**Figure 4. Analysis of the spectral response of the plasmonic crystal based photodetectors.** Simulated Absorption in Au with varying lattice parameter of a structure with r = 150nm (a) considering the complex refractive index of ITO, (b) real part only refractive index of ITO, (c) $TiO_2$ – Au corrugated interface and (d) A dielectric ($TiO_2$ - ITO) array on top of a flat Au film. Vertical dashed lines indicate two of the lattice parameters employed to fabricate the devices from Fig 2. The white dots represent the points where the field profiles are shown in Fig. 3. Modes at the metal can be excited either by coupling to SPPs with a dielectric grating (panel d), a metallic grating (panel c) and the resulting interaction of both (panel b).

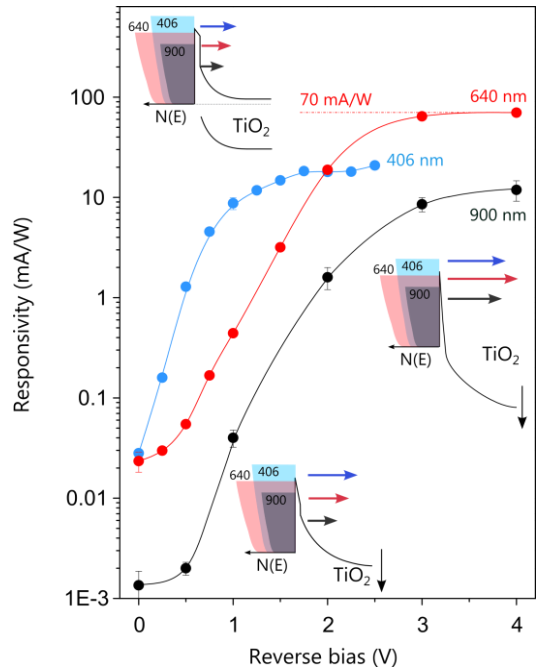

**Figure 5. Device optoelectronical performance.** Responsivity as a function of applied reverse bias for a representative device (*r*=150nm, *L*=450nm) at three different illumination wavelengths. Negatively biasing the junction results in a prominent increase of responsivity as a higher fraction of the hot electron population is able to tunnel through the $Al_2O_3/TiO_2$ barrier. Depending on the illumination energy different responsivity onsets and saturation voltages are obtained, as the tunneling probability increases for lower energies with increasing bias. A maximum R = 70 mA/W (yielding an estimated IQE of 30%, see suppl. section S5) is obtained for 640 nm illumination at 4V reverse bias.

# Supplementary Information

**Multiband Tunable Large Area Hot Carrier Plasmonic-Crystal Photodetectors**

F. Pelayo García de Arquer[†], Agustín Mihi[†,*], Gerasimos Konstantatos*

*<gerasimos.konstantatos@icfo.es>, agustin.mihi@icfo.es

† equal contribution

ICFO - Institut de Ciències Fotòniques, Mediterranean Technology Park 08860 Castelldefels, Barcelona, Spain

**Contents:**

**S1. Plasmonic crystal spectral comparison with a flat reference**

**S2. Absorption variation with coating thickness and refractive index**

**S3. Absorption variation with pillar radius and depth**

**S4. Dark current- voltage characteristics**

**S5. IQE estimation**

**S6. Photocurrent as a function of incident power**

**S7. Devices temporal response**

## S1. Plasmonic crystal spectral comparison with a flat reference

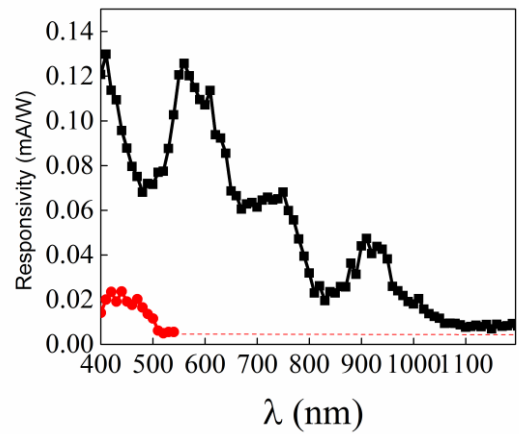

**Figure S1.** Experimental responsivity of a plasmonic crystal photodetector with r=150 nm and L=450 nm (black dotted line) compared with a flat reference (red dotted line) where only Au interband transitions contribute. Measurements in the flat device could only be acquired until 550nm, after that the response was below the detection limit of the set up.

## S2. Absorption variation with coating thickness and refractive index

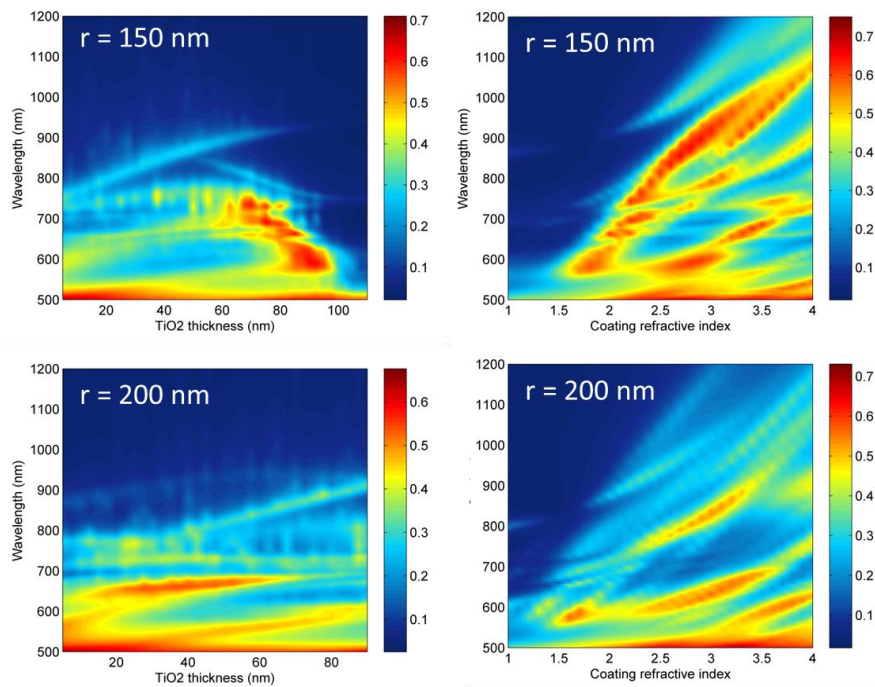

**Figure S2.** Simulated Absorption in the Au as a function of semiconducting coating thickness (top left) and and refractive index (top right) for an imprinted cylinder radii r = 150 nm (top) and r = 200 nm (bottom). $TiO_2$ layer thickness limits the metal cylinder radius and modifies the position and intensity of propagating and localized resonances. The plasmon resonances are very sensitive to coating refractive index (right column), showcasing the importance of a proper semiconductor election for hot-electron device applications.

## S3. Absorption dependence with pillar radius and depth

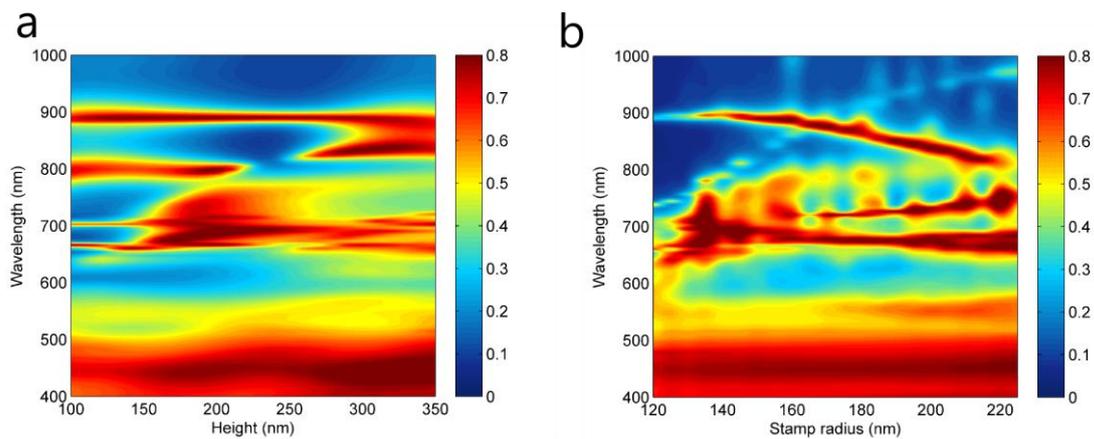

**Figure S3.** Variation of the simulated absorption in Au as a function of imprinted pillar height (a) and radius (b) for $L = 450$ nm and $t = 60$ nm. It can be seen that the resonance at 900 nm does not depend on pillar height, whereas other features change with this parameter, such as the resonances at 700 nm and 800 nm, which is in agreement with the proposed mode description in the main text. The behaviour with the initial air cylinder radius (b) is more complex, as the filling fraction, and therefore the effective refractive index of the gratings, also change with this parameter. The variation of the cylinder radius results in a slight blue-shift of the 900 nm resonance and position and intensity change for the resonances in the 600-800 nm window. These changes are attributed to the resulting different metallic pillars (as the ITO and $TiO_2$ thicknesses were kept constant) obtained by increasing the diameter of the initial cylinders.

## S4. Dark current- voltage characteristics

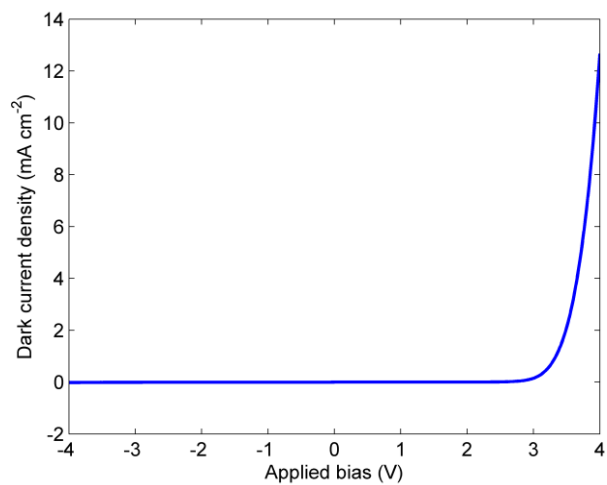

**Figure S4.** Dark current voltage characteristics of a representative device depicting the Schottky junction established between the Au and the $TiO_2$ substrate.

## S5. Estimation of absorption for IQE calculation

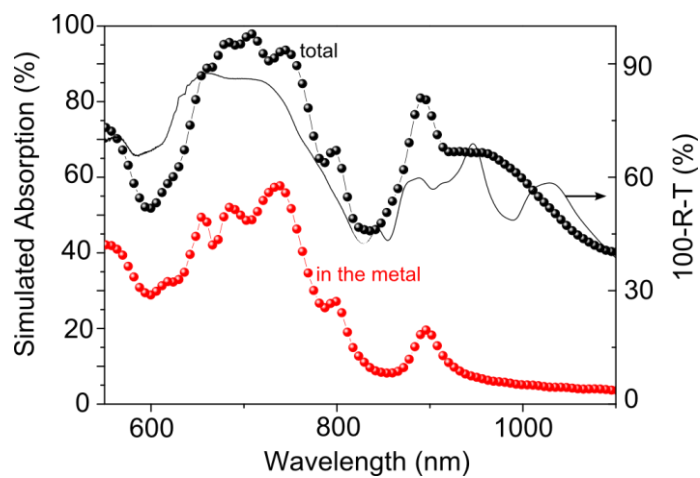

**Figure S5.** Simulated Absorption in the photodetector architecture (calculated as 100-R, black dotted line) versus simulated Absorption in the gold layer (red dotted line). The value of 100-R measured in an actual device is presented in solid line for comparison (right axis).

The Reflectance was measured using an Agilent 660 FTIR attached to a microscope with a 4X objective (NA 0.1) and a spatial mask (solid black line). This experimental curve was reproduced with FDTD simulations as described in the main text. Once the total absorption (black dotted line) was simulated, only the absorption in the metal electrode (red dotted line) was considered to calculate the IQE. Considering the obtained EQE of 12%, we estimated a 30% IQE if only the simulated absorption in the metal is considered (red dotted line) disregarding the amount of light absorbed by the ITO or trapped within the device thickness; if the total experimental absorption is considered instead (100-R, solid line in right axis) in the calculation, a lower bound of 17% IQE is obtained.

Au refractive index was fitted from Johnson and Christy data. The absorbing ITO was modeled as follows. The absorption coefficient was calculated from the experimental absorption of films of known thicknesses, measured with a Cary-Varian 5000 spectrophotometer equipped with an integrating sphere attachment. The absorption coefficient allows the straightforward calculation of the refractive index imaginary part ($\alpha = 4\pi\kappa/\lambda$). The complex function $n = n' + i\kappa$ is analytical and therefore has to satisfy Kramers-Kronig equations. This allows the determination of the complex refractive index which is used along the simulations.

## S6. Photocurrent as a function of incident power

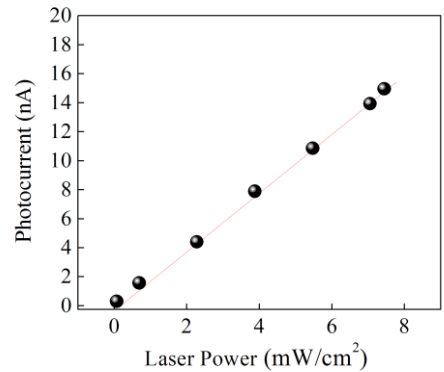

**Figure S6.** Photocurrent (at short circuit conditions) as a function of incident laser power for 635 nm illumination illustrates the dynamic properties of our devices.

## S7. Temporal response

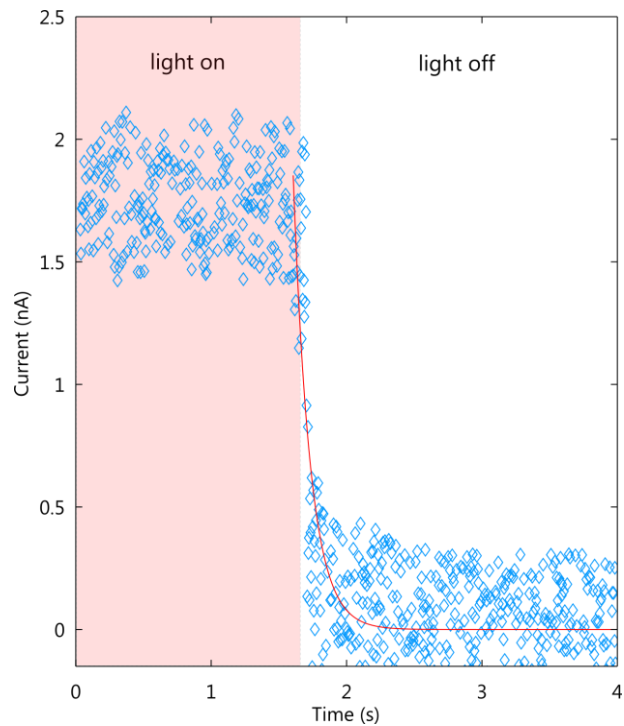

**Figure S7.** Time trace of a representative device at short-circuit conditions. The decay is fitted with a single 125 ms exponential, in excellent agreement with the extracted RC constant of the device (the measured $C = 5$nF and $R = 25$MΩ at the same conditions yield a $\tau = RC$ of 125 ms).